\documentclass[11pt]{article}
\usepackage{blois,epsfig,amssymb,amsmath}

\def\Journal#1#2#3#4{{#1} {\bf #2}, #3 (#4)}

\def\PLB{{\em Phys. Lett.}  B}
\def\PRL{\em Phys. Rev. Lett.}
\def\PRD{{\em Phys. Rev.} D}

\def\be{\begin{equation}}
\def\ee{\end{equation}}
\def\bea{\begin{eqnarray}}
\def\eea{\end{eqnarray}}

\begin{document}
\vspace*{4cm}
\title{MODEL-INDEPENDENT ANALYSIS OF FORWARD-BACKWARD 
ASYMMETRY OF TOP QUARK PRODUCTION AT THE TEVATRON}

\author{DONG-WON JUNG}
\address{Department of Physics, National Tsing Hua University, 
\\ Hsinchu, Taiwan 300}

\author{ P. KO}
\address{School of Physics, Korea Institute for Advanced Study, \\
Seoul 130-722, Korea}

\author{JAE SIK LEE}
\address{Physics Division, National Center for Theoretical Sciences, 
\\ Hsinchu, Taiwan 300}

\author{SOO-HYEON NAM}
\address{Korea Institute of Science and Technology Information, 
\\ Daejeon 305-806, Korea}

\maketitle\abstracts{
We perform a model independent analysis on 
$q \bar{q} \rightarrow t \bar{t}$ using an effective 
lagrangian with dim-6 four-quark operators, and derive  
necessary conditions on new physics that are consistent with 
the $t\bar{t}$ production cross section 
and the forward-backward (FB) asymmetry ($A_{\rm FB}$) 
measured at the Tevatron.  
We also propose a new FB spin-spin correlation that is strongly 
correlated with the $A_{\rm FB}$, and   
discuss possible new physics scenarios that 
could generate such dim-6 operators.  
}

\section{Introduction}

The $A_{\rm FB}$ of the top quark is one of the interesting observables 
related with top quark.  Within the Standard Model (SM), 
this asymmetry vanishes at leading order in  QCD because of $C$ symmetry. 
At next-to-leading order [$O(\alpha_s^3)$], 
a nonzero $A_{\rm FB}$  can develop from the interference
between the Born amplitude and two-gluon intermediate state, 
as well as the gluon bremsstrahlung and gluon-(anti)quark scattering 
into $t \bar{t}$, 
with the prediction $A_{\rm FB}\sim 0.078$ \cite{Antunano:2007da}.  
The measured asymmetry has been off the SM prediction by $2 \sigma$ 
for the last few years, albeit a large experimental uncertainties. 
The  measurement in the $t\bar{t}$ rest frame before this meeting was 
\cite{cdf2009}
\begin{equation}
A_{\rm FB} \equiv \frac{N_t ( \cos\theta \geq 0) - N_{\bar{t}} 
( \cos\theta \geq 0 )}{N_t ( \cos\theta \geq 0) + N_{\bar{t}} 
( \cos\theta \geq 0 )} =  (0.24 \pm 0.13 \pm 0.04) 
\end{equation}
with $\theta$ being the polar angle of the top quark with 
respect to the incoming proton in the $t\bar{t}$ rest frame.
This $\sim 2\sigma$ deviation stimulated some speculations on new physics
scenarios \cite{Choudhury:2007ux,Djouadi:2009nb,Ferrario:2009bz,Jung:2009jz,ko}. 

On the other hand, search for a new resonance decaying into  
$t\bar{t}$ pair has been  carried out  at the Tevatron. 
As of now, there is no clear signal for such a new resonance \cite{cdf2009}.  
Therefore, in this talk, I assume that  a new physics scale relevant to 
$A_{\rm FB}$ is large enough so that  production of a new particle is 
beyond the reach of the Tevatron \cite{ko}, 
which makes a key difference between 
our work and other literatures on this subject 
\cite{Choudhury:2007ux,Djouadi:2009nb,Ferrario:2009bz,Jung:2009jz}.  
Then it is adequate to integrate out the heavy fields, and use the 
resulting 
effective lagrangian approach in order to study 
new physics effects on $\sigma_{t\bar{t}}$ and $A_{\rm FB}$.  
At the Tevatron, the $t\bar{t}$ production is dominated by $q\bar{q} 
\rightarrow t\bar{t}$, and it would be sufficient to consider dimension-6
four-quark operators (the so-called contact interaction terms) 
to describe the new physics effects on the $t\bar{t}$ production 
at the Tevatron.  A similar approach was adopted for the dijet 
production to constrain the composite scale of light quarks, 
and we are proposing the same analysis for the $t\bar{t}$ system.  

\section{Model independent analysis}
\subsection{Lagrangian}

Our starting point is the effective lagrangian with dimension-6 
operators relevant to the $t\bar{t}$ production at the Tevatron: 
\begin{equation}
\mathcal{L}_6 = \frac{g_s^2}{\Lambda^2}\sum_{A,B}
\left[C^{AB}_{1q}(\bar{q}_A\gamma_\mu  q_A)(\bar{t}_B\gamma^\mu t_B)  + 
C^{AB}_{8q}(\bar{q}_A T^a\gamma_\mu q_A)(\bar{t}_B T^a\gamma^\mu  t_B)\right]
\end{equation}
where $T^a = \lambda^a /2$, $\{A,B\}=\{L,R\}$, and 
$L,R \equiv (1 \mp \gamma_5)/2$ 
with $q=(u,d)^T,(c,s)^T$.  
%
Using this effective lagrangian, we calculate the cross section up to 
$O(1/\Lambda^2)$, keeping only the interference term between 
the SM and new physics contributions.  
The above effective lagrangian was also discussed in 
Ref.~\cite{Hill:1993hs}, where the $t$ quark was treated as 
$SU(2)_L \times SU(2)_R$ singlet and top currents were 
decomposed into vector and axial vector currents, rather than
chirality basis as in our case.   

\begin{figure*}
\includegraphics[width=7cm]{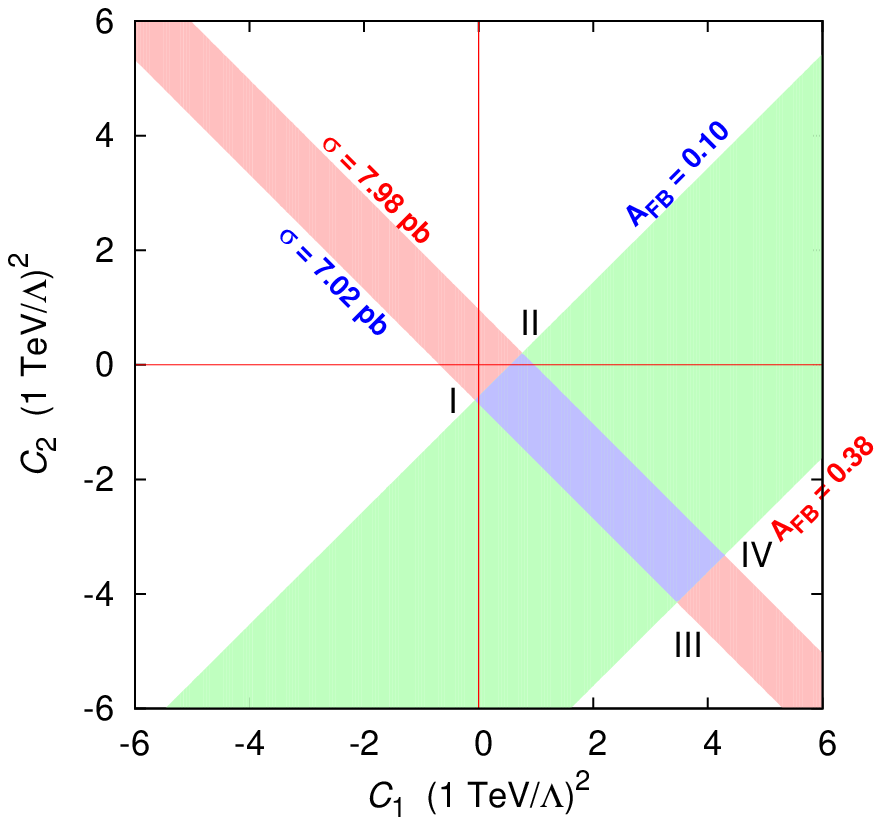}
\includegraphics[width=7cm]{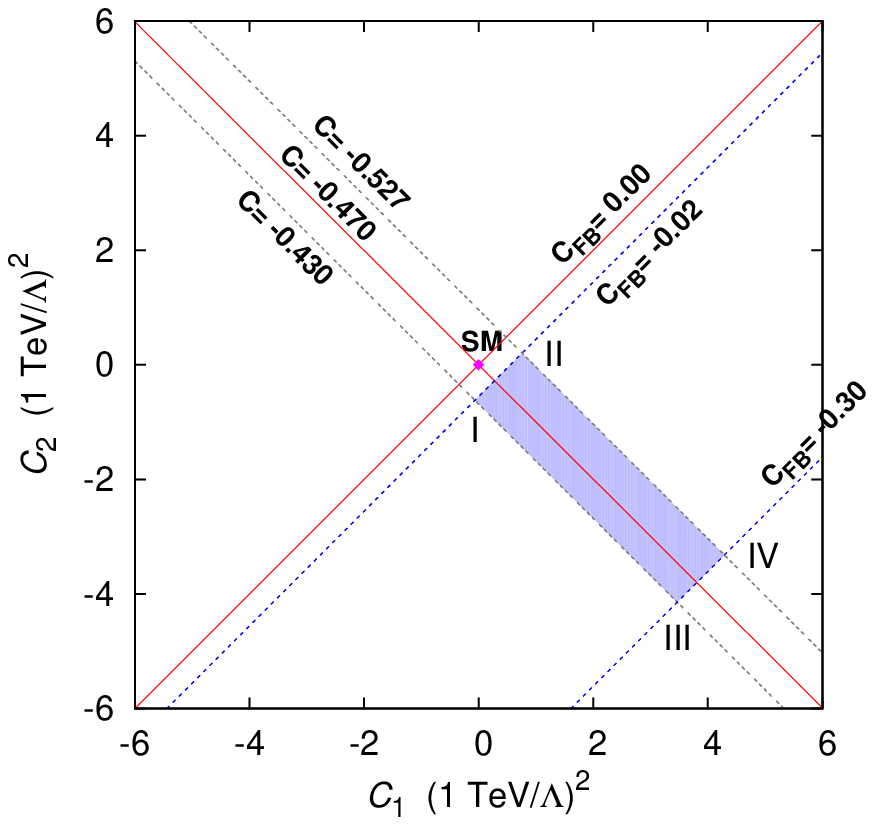} \\%
\caption{\label{} (a) The region in $( C_1 , C_2 )$ plane that is
consistent with the Tevatron data  at the $1 \sigma$ level:   
$\sigma_{t\bar{t}} = (7.50 \pm 0.48)$ pb  
and $A_{\rm FB} = (0.24 \pm 0.13 \pm 0.04)$.
(b) the spin-spin correlations $C$ and $C_{FB}$.
}
\label{fig:1a}
\end{figure*}

\subsection{Origin of FB Asymmetry}

It is straightforward to calculate the amplitude for 
$q (p_1) + \bar{q} (p_2) \rightarrow t (p_3) + \bar{t} (p_4)$ 
using the above effective lagrangian and the SM. 
The squared amplitude summed (averaged) over the final (initial) 
spins and colors is given by
\begin{eqnarray}
\overline{|{\cal M}|^2} 
& \simeq  & \frac{4\,g_s^4}{9\,\hat{s}^2} \left\{
2 m_t^2 \hat{s} \left[
1+\frac{\hat{s}}{2\Lambda^2}\,(C_1+C_2)
\right] s_{\hat\theta}^2   \right. 
\\ 
& + & \left. 
\frac{\hat{s}^2}{2}\left[ \left(1+\frac{\hat{s}}{2\Lambda^2}\,(C_1+C_2)\right)
(1+c_{\hat\theta}^2)
+\hat\beta_t\left(\frac{\hat{s}}{\Lambda^2}\,(C_1-C_2)\right)c_{\hat\theta}
\right]\right\}   \nonumber 
\label{eq:ampsq}
\end{eqnarray}
where $\hat{s} = (p_1 + p_2)^2$, $\hat\beta_t^2=1-4m_t^2/\hat{s}$,
and $s_{\hat\theta}\equiv \sin\hat\theta$ and 
$c_{\hat\theta}\equiv \cos\hat\theta$ 
with $\hat{\theta}$ being the polar
angle between the incoming quark and the outgoing top quark in the 
$t\bar{t}$ rest frame. 
And the couplings are defined as:
$C_1 \equiv C_{8q}^{LL}+C_{8q}^{RR}$ and 
$C_2 \equiv C_{8q}^{LR}+C_{8q}^{RL}$. 
Since we have kept only up to the interference terms, there are 
no contributions from  the color-singlet operators with coupling 
$C_{1q}^{AB}$. 
The term linear in $\cos\hat{\theta}$ could
generate the forward-backward asymmetry 
which is proportional to $\Delta C \equiv (C_1 - C_2)$.
Note that both light quark and top quark should have chiral couplings
to the new physics in order to generate $A_{\rm FB}$ at the tree level
(namely $\Delta C \neq 0$).  This parity violation, if large, 
could be observed in the nonzero (anti)top spin polarization \cite{progress}. 

In Fig.~\ref{fig:1a}, we show the allowed region in the $(C_1,C_2)$ plane 
that is consistent with the Tevatron data  at the $1 \sigma$ level.
The allowed region is around $0 \lesssim C_1 \lesssim 4$ and 
$-4 \lesssim  C_2 \lesssim + 0.5$. The negative sign of $C_2$ is preferred
at the 1 $\sigma$ level.   

\subsection{A New Spin-spin Correlation}

Another interesting observable which is sensitive to the chiral 
structure of new physics affecting $q\bar{q} \rightarrow t\bar{t}$ 
is the top quark spin-spin correlation \cite{Stelzer06,progress}:
\begin{equation}
C = \frac{\sigma(t_L\bar{t}_L + t_R\bar{t}_R) - 
\sigma(t_L\bar{t}_R + t_R\bar{t}_L)}{\sigma(t_L\bar{t}_L + t_R\bar{t}_R) + 
\sigma(t_L\bar{t}_R + t_R\bar{t}_L)} \,.
\end{equation}
%
Since new physics must have chiral couplings both to light quarks and 
top quark, the spin-spin correlation defined above will be affected.
From Eq. (\ref{eq:ampsq}),  it is clear the spin-spin correlation 
Eq.~(4) is sensitive to $(C_1 + C_2)$, 
since the linear term in $\cos\hat{\theta}$ does not contribute to
the correlation $C$ after integration over $\cos\hat{\theta}$.
On the other hand, if one considers the forward and the backward regions 
separately,  the spin-spin correlation would depend on $( C_1 - C_2 )$
and will be closely correlated with $A_{\rm FB}$. 
Therefore we propose a new spin-spin FB asymmetry $C_{FB}$  defined as 
\begin{equation}
C_{FB} \equiv  C (\cos\theta \geq 0) -  C (\cos\theta \leq 0)  ,
\end{equation}
where $C(\cos\theta \geq 0 (\leq 0))$ implies the cross sections in the 
numerator of Eq.~(4) are obtained for the forward (backward) region: 
$\cos\theta \geq 0 (\leq 0)$.
In Fig.~\ref{fig:1a} (b),  we show the contour plots for the $C$ and $C_{FB}$ 
in the $(C_1 , C_2 )$ plane along with the SM prediction at LO.
There is a clear correlation between $C_{FB}$ and $A_{FB}$ in Fig.~\ref{fig:1a}, 
which must be observed in the future measurements 
if the $A_{\rm FB}$ anomaly is real and  
a new particle is too heavy to be produced at the Tevatron.  

\section{Explicit Models}
So far, we considered dim-6 four-quark operators that could
affect the $t\bar{t}$ productions at the Tevatron, and found 
the necessary conditions for accommodating $A_{\rm FB}$.
In Ref.~\cite{ko}, we also considered the explict models with new 
particles with various spins and colors that could 
affect $A_{\rm FB}$. 
In Table~\ref{tab:newparticles}, we show the new particle exchanges  
under consideration and the signs of the couplings $C_1, C_2$ induced by them.
We found that the four types of exchanges of $V_8$, $\tilde{V}_8$,
$\tilde{S}_1$, and $S_{13}^{\alpha\beta}$ could give rise to the large positive 
$A_{\rm FB}$ at the 1-$\sigma$ level.
It would be interesting to search for new vector or 
scalar particles that satisfy the above conditions at LHC.
For more quantitative discussions,  we have to study the full amplitude 
without integrating out new heavy particles, 
the detailed study of which will be presented in the 
future work \cite{progress}.

\begin{table}[t]
\caption{\label{tab:newparticles}
{\it
New particle exchanges and the signs of induced couplings $C_1$ and $C_2$}
}
\begin{center}
\begin{tabular}{l|c|c|c|c}
\hline\hline
& & & & \\[-0.3cm]
New particles & couplings & $C_1$ & $C_2$ & 1 $\sigma$ favor \\[0.1cm]
\hline\hline
& & & & \\[-0.3cm]
$V_8$ (spin-1 FC octet) & $g^{L,R}_{\,8q,8t}$ & indef. & indef. & $\surd$ \\[0.1cm]
$\tilde{V}_1$ (spin-1 FV singlet) & $\tilde{g}^{L,R}_{1q}$ & $-$ & $0$ & $\times$ \\[0.1cm]
$\tilde{V}_8$ (spin-1 FV octet) & $\tilde{g}^{L,R}_{8q}$ & $+$ & $0$ & $\surd$ \\[0.1cm]
\hline
& & & & \\[-0.2cm]
$\tilde{S}_1$ ~~(spin-0 FV singlet) & $\tilde\eta^{L,R}_{1q}$ & $0$ & $-$ & $\surd$ \\[0.1cm]
$\tilde{S}_8$ ~~(spin-0 FV octet) & $\tilde\eta^{L,R}_{8q}$ & $0$ & $+$ & $\times$ \\[0.1cm]
$S_2^\alpha$ ~\,(spin-0 FV triplet) & $\eta_{3}$ & $-$ & $0$ & $\times$ \\[0.1cm]
$S_{13}^{\alpha\beta}$ \,(spin-0 FV sextet) & $\eta_{6}$ & $+$ & $0$ & $\surd$ \\[0.1cm]
\hline\hline
\end{tabular}
\end{center}
\end{table}

\section{Conclusions}

In this talk,  I presented a model independent study
of $t\bar{t}$ production cross section and $A_{\rm FB}$ 
at the Tevatron using dimension-6 contact interactions.
We derived conditions for the couplings
of four-quark operators that could generate the FB asymmetry 
observed at the Tevatron [Fig.~\ref{fig:1a}]. 
Then we considered the $s-$, $t-$ and $u-$channel exchanges of 
spin-0 and spin-1 particles whose color quantum number is 
either singlet, octet, triplet or sextet.   
Our results in Fig.~1 and Table 1 
encode the necessary conditions for the underlying new physics 
in a compact and an effective way, when those new particles
are too heavy to be produced at the Tevatron but still affect $A_{\rm FB}$.  
If these new particles could be produced directly at the Tevatron or 
at the LHC,  we cannot use the effective lagrangian any more. 
We have to study specific models case by case, and anticipate 
rich phenomenology at colliders as well as at low energy. 
Detailed study of these issues 
will be discussed in the future publications \cite{progress}. 

\section*{Acknowledgments}
This work was supported in part by Korea Neutrino Research Center (KNRC) 
through National Research Foundation of Korea Grant.

\end{document}